\documentstyle[11pt]{article}
\input epsf
\begin{document}
\def\dx{\partial_x}
\def\deltamn{\delta_{m+n,0}}
\def\deltaxy{\delta(x-y)}
\def\levi{\epsilon_{ijk}}

\def\rlx{\relax\leavevmode}
\def\inbar{\vrule height1.5ex width.4pt depth0pt}
\def\IZ{\rlx\hbox{\small \sf Z\kern-.4em Z}}
\def\IR{\rlx\hbox{\rm I\kern-.18em R}}
\def\ID{\rlx\hbox{\rm I\kern-.18em D}}
\def\IC{\rlx\hbox{\,$\inbar\kern-.3em{\rm C}$}}
\def\IN{\rlx\hbox{\rm I\kern-.18em N}}
\def\one{\hbox{{1}\kern-.25em\hbox{l}}}
\def\smallfrac#1#2{\mbox{\small $\frac{#1}{#2}$}}

\begin{titlepage}

Revised, November 1995 \hfill{UTAS-PHYS-95-14}\\
\mbox{{\it Class Q Grav} (to appear)}\hfill{hep-th/yymmdd}
\vskip 1.6in
\begin{center}
{\Large {\bf Modified Relativity from the $\kappa$-deformed Poincar\'{e}
Algebra}}
\\[5pt]
\end{center}

\normalsize
\vskip .4in

\begin{center}
J P Bowes \hspace{3pt}
and \hspace{3pt} P D Jarvis
\par \vskip .1in \noindent
{\it Department of Physics, University of Tasmania}\\
{\it GPO Box 252C Hobart, Australia 7001}
\end{center}
\par \vskip .3in

\begin{center}
{\Large {\bf Abstract}}\\
\end{center}

\vspace{1cm}

The theory of the $\kappa$-deformed Poincar\'{e} algebra is applied to
the analysis of various phenomena in special relativity,
quantum mechanics and field theory.
The method relies on the development of
series expansions in $\kappa^{-1}$ of the generalised
Lorentz transformations, about the special-relativistic limit.
Emphasis is placed on the underlying assumptions needed
in each part of the discussion, and on \emph{in principle} limits
for the deformation parameter, rather than on rigorous
numerical bounds.
In the case of the relativistic
Doppler effect, and the Michelson-Morley experiment, comparisons with
recent experimental tests yield the relatively weak lower
bounds on $\kappa c$ of 90 $eV$ and 250 $keV$, respectively.
Corrections to the Casimir effect and the Thomas precession are also discussed.

\end{titlepage}

\section{Introduction and Main Results}

In recent years there has been much interest in the implications for physical
theories of
the notion of `deformed' symmetries. First introduced in the context of the
inverse
scattering transform and exactly solvable models in statistical mechanics,
the study of
`quantum groups' [1] in particular entrains the mathematical structures of Hopf
algebras, which provide a more powerful and general language for the
discussion of
symmetry principles than the traditional one of Lie algebras and Lie
groups. The
deformations also typically involve one or more scalar parameters; these may be
coupling constants as in the case of lattice systems, but in the context of
deformed
theories generally, may be new types of fundamental dimensionful constants.
It is
natural to speculate on their role as, say, cutoff parameters in the
renormalization of
quantum field theory, or unification scales in applications to symmetry
breaking
phenomena.

The present work addresses the question of deformed space-time symmetries. In
particular we take up the case of the so-called $\kappa$-deformed
Poincar\'{e} algebra introduced
recently and studied in several papers [2]. In this case the deformation
parameter gives
a fundamental inverse length or energy scale beyond which deviations from
special
relativity are expected. The paper concerns the detailed working-out of the
implications
of this class of deformed space-time algebra for a variety of physical,
experimentally
tested effects. Specifically, we develop a series expansion in inverse
powers of
$\kappa$ for the
generalised `Lorentz transformations' about the special relativistic limit,
and we
consider the lowest order corrections to the experimental predictions of
special relativity
for the various effects, in order to deduce lower limits on $\kappa$.
Our main results (see
\S 2 below) are that the existing data on the transverse Doppler shift and the
Michelson-Morley experiment yield limits of 90 $eV$ and 250 $keV$
 respectively for $\kappa c$.

The $\kappa$-deformed space-time symmetries [3] were first derived from a
systematic
application of the formalism of $q$-deformations to real forms of simple
Lie algebras
related to compactified Minkowski space, followed by appropriate
contraction [3,4].
Since their introduction further developments have been considered,
including higher
dimensional [5,6] and supersymmetric [7,8,9] versions, the formalism of
$\kappa$-relativistic
fields [10], the integration of the infinitesimal transformations to finite
transformations
[11], $\kappa$-Dirac and higher spin equations [11,12], as well as the
elements of induced
representation theory [13]. Formal developments have included consideration
of the
coproduct structure and tensor operators, [14] for the $\kappa$-deformed
algebra. Analogously
to the standard theory of quantum groups, theorems have been proved [14] which
ensure that for generic values of $\kappa$,
 representations of the undeformed algebra also
admit actions under the deformed algebra.

Deformations of symmetries have also arisen in a more geometrical arena in
connection
with `non commutative geometry' [15]. In this dual approach, abstract
generalisations of
functions on manifolds would imply the existence of coordinates which are
no longer
point like, but belong to a non commutative algebra. Applied to space-time
symmetries
this leads to consideration of the quantum group $SL_q(2,\IC)$ [16]. This
framework seems
a natural one for the introduction of discrete substructure, as might be
expected in
quantum theories of gravity at the highest Planck energies. Indeed one proposal
involving applications of Hopf algebras in geometric quantisation leads to
a natural
`bicrossproduct' structure [17,18] in which the role of Planck's constant
in quantisation
of the Weyl algebra of quantum mechanics is dually matched by a second
fundamental
deformation parameter in the quantisation and coproduct structure of
accompanying
symmetries. In particular for model systems [17,18] this parameter can be
plausibly
related to Newton's gravitational constant. The further ramifications of
this line of
thinking lead to a fertile concept of `braided' generalisations of Lie
algebras [19].

It has recently been shown that the $\kappa$-deformed algebra itself admits
a bicrossproduct
structure in the above sense [20], giving some indication that the
numerical value of the
deformation parameter $\kappa$ might be comparable with the Planck energy.
At the same
time, this suggests that the correct context for detailed study of the physical
consequences of the $\kappa$-relativity should be that of noncommutative
geometry [20].

In the present work we eschew such theoretical sophistication, preferring
to work with
the deformed relativity as a theory which induces small but testable
corrections to
experimental results: we claim that, even with the full apparatus of non-commutative
geometry, if indeed it is required to make a consistent theory, actual
deviations from
relativity will involve much the same type of algebraic consequences as
given here.
Here we treat some of the main predictions of special relativity; a more
comprehensive
analysis would consider the whole gamut of Post-Newtonian
parametrisations [21] (see below).

For the purposes of this work the detailed Hopf structure of the
$\kappa$-Poincar\'{e} algebra is
not essential (the salient definitions are given in the appendix for
completeness).
Amongst the Lorentz generators, deformed commutation relations arise only
for the
boosts $L_i$ for $i=1,2,3$, and between the boosts and the momentum
generators $P_0$, and $P_i$,
with the standard expressions modified by rational functions of $\exp{P_0/
\kappa}$;
commutation
relations for the angular momentum generators $M_i$ are undeformed. The
quadratic
Casimir invariant becomes
\begin{equation}
	C_1 = {\mathbf{P}}^2 - (2 \kappa \sinh \frac{P_0}{2 \kappa})^2
	\label{dispersion}
\end{equation}
with an expression for the spin invariant corresponding to a Pauli-Lubanski
vector
which has been deformed in an analogous way.

The modified dispersion relation (\ref{dispersion}) leads directly to
experimental determinations of
(limits on) $\kappa$, and high energy astrophysical processes have been
analysed by assuming
that photons are `massless', and that (\ref{dispersion}) prescribes an
effective energy dependence for
the photon phase velocity $\omega /k=E/p$. Thus from the coincidence
measurements on the
arrival times of sharp bursts from distant events over widely different
energies [22] the
lower limit on $\kappa$  of $10^{12} \: GeV$ was derived.

Although (\ref{dispersion}) entails a mild deviation from special relativity at
laboratory energies, in
the context of quantum field theory it is possible that the
entire ultraviolet structure is
changed. The simplest manifestation of this is the
Casimir effect, which depends only
on the zero point energy of virtual modes.
Following the standard text book derivations
[23,24] for plane square parallel plates of side $L$ and separation $d$,
the $\kappa$--relativistic
dispersion relation leads to the usual Casimir interaction energy plus a
series of $\kappa$--dependent
corrections
\begin{equation}
	U(d) = -\frac{\pi^2\hbar c L^2}{720 d^3}
	\left( 1+ \frac{1}{84} \left(\frac{\pi \hbar}{\kappa d}\right)^2
	+ \frac{9}{896} \left(\frac{\pi \hbar}{\kappa d}\right)^4 + \cdots \right)
	\label{Casimireffect}
\end{equation}
with the general $n$th coefficient (for $n>1)$ being given in terms of
Bernoulli numbers, [25]         	
\begin{equation}
	U(d) \cdot \frac {2 d^3}{\pi^2\hbar c L^2} = (-1)^{(n+1)}\frac{B_{(2n+2)}}
	{(2n+1)(2n+2)(2n-1)}\frac{\Pi_{m=1}^{n-1}(2m-1)}{\Pi_{m=1}^{n-1} 2m}
	\left(\frac{\pi \hbar}{2\kappa d}\right)^{2(n-1)}.
	\label{Bernoulli}
\end{equation}
Due to the $n ! $ behaviour of the Bernoulli numbers for large $n$,
(\ref{Casimireffect})
is in fact divergent for all finite
$\kappa$ and $d$, which is of course very different from the case of the
usual Casimir
effect. On the other hand, if we assume that (\ref{Casimireffect}) is an
asymptotic
series, as often arises in field theory, it is found that the
Casimir force will be $\kappa$-affected at small separations only.
In practice this is a very difficult
experimental regime (see for example [25]) due to the skin depth effect in
real conductors, so that any
bound on $\kappa c$ from existing data is in the $eV$ range at best.
However, the effect of the
deformation has potentially drastic consequences for a variety of phenomena
which
involve the Casimir effect, such as black hole evaporation.

Before turning to the analysis of
special relativistic tests, we present
a further example, that of the Thomas precession, to illustrate our general
approach.
Whereas the above
discussion of the Casimir effect relies essentially on the modifications
to the energy-momentum relationship, quantum mechanical wave equations in
$\kappa$--relativity, especially for
spinning particles (see for example [11], [12]) necessarily must address
issues in the representation theory of the full deformed algebra. On the
other hand, the original arguments for the relativistic
origins of the atomic spin-orbit coupling may be discussed
using purely kinematical arguments involving, for example, the commutator
of two boosts in perpendicular directions. The same textbook argument goes
through
in the $\kappa$-deformed algebra, and would lead to a modified spin-orbit
coupling of the type
\begin{equation}
	\Delta E = \frac{1}{r}\frac{dV(r)}{dr}
	\left(\frac{\Sigma_\kappa}{2 m^2 c^2}\right) \mathbf{S \cdot L},
	\label{Thomas}
\end{equation}
where $\Sigma_\kappa = 1 + O(m^2 c^2 /\kappa^{2})$ (see below for comments
about expansions in powers of $\kappa$).
This semi-classical derivation is obviously incomplete, in
that it has not required $\kappa$--deformed Dirac or Maxwell equations.
However, it still gives
some insight into how the deformation will alter the energy levels of the
hydrogen
atom, and moreover because of the paucity of underlying assumptions,
we would claim that \emph{any} consistent theory must
contain terms of the same general form.

In \S 2 below we return to the question of special relativistic
tests of $\kappa$-relativity.
The derivations use the results of the appendix, where we present the
evaluation of the
series expansion of the `$\kappa$-deformed Lorentz transformations' about
the special
relativistic limit in inverse powers of $\kappa$.
This involves various manipulations of the
Jacobi elliptic functions which arise from integrating the
infinitesimal transformations
generated by the $\kappa$-deformed boosts. The result is that the usual
linear momentum transformation $p_\mu' = {\Lambda_\mu}^\nu p_\nu$
is now the first term in the expansion, the general relation being of the
form
$p_\mu' = L_\mu(p)$. In \S 2 these expansions are used in an
analysis of two fundamental experimental tests of relativity: the
transverse Doppler
effect, and the Michelson-Morley experiment.
The former is a direct application of the
formulae of the appendix. The latter is analysed by
modelling radiation between the
etalon plates as a scalar field. However, in order to complete the discussion,
additional assumptions about coordinate space aspects of
$\kappa$-relativity, and also about the superposition principle, are required.
These are discussed as the analysis is developed.

In our conclusions (\S 3), we summarize our results and give some
final remarks on the prospects for a more comprehensive, systematic
analysis of experimental data which may provide tests of deformed relativity.
Our work is based on [26] (unpublished), but the
presentation of the material given in this paper is in principle
self-contained.

\section{Special relativistic tests}

\subsection*{Transverse Doppler Effect}

A fundamental relativistic phenomenon is the relativistic Doppler effect,
which has in the past been used as an important experimental test of special
relativity. We can obtain the $\kappa$-generalised relativistic Doppler
effect from the $\kappa$ deformed four-momentum transformations
(\ref{p0transf})-(\ref{p3transf}) by simply replacing
the four-momentum parameters by their corresponding four wave-numbers, using
the deBroglie relation.
One of the main problems associated with experimental tests of the
relativistic Doppler
effect is the presence of two distinct contributions to the total
Doppler effect observed. These contributions are best described as the
`relativistic component' which we are primarily interested in, and the
`non-relativistic contribution', {\it ie} $k_{0}'=k_{0}(1+v/c)$. In the
transverse
Doppler effect, this relativistic component is the sole contributor, {\it ie}
\begin{equation}
\frac{k_{0}'}{k_{0}}=\cosh\alpha \equiv \gamma,
\end{equation}
which is found to $\kappa$ generalise to,
\begin{equation}
\frac{k_{0}'}{k_{0}}=\cosh\alpha+\frac{\hbar^{2}k_{0}^{2}}{8\kappa^{2}}
\left[\frac{\cosh\alpha}{3}-\frac{\cosh^{3}\alpha}{3}+ \frac{\alpha\sinh
\alpha}{2}+\frac{\sinh^2\alpha \cosh \alpha}{2}\right]
+0(\kappa^{-4})
\end{equation}
(compare with (\ref{p0transf})-(\ref{p3transf})). In the past experimental
tests of the
relativistic Doppler effect have concentrated on the measurement of this
transverse component. There are however practical problems associated with this
measurement,
due to the fact that any small aberrant longitudinal component will swamp this
weak transverse effect.

Recent measurements of the relativistic Doppler
effect have however been improved (see for example [27]) by using a technique in
which the relativistic component is isolated using
a two-photon absorption process (TPA). In two-photon spectroscopy, the
first order Doppler shift is effectively absent, and the second order term
becomes
dominant. By measuring the frequency difference between a two-photon
transition in a fast neon-atom beam and a stationary neon sample, using
two lasers, the second order Doppler
effect was confirmed [27] to accuracy $4\times 10^{-5}$.

The analysis of this experiment can be performed
using our $\kappa$-modified expressions for the
relativistic Doppler effect, obtained explicitly from
(\ref{p0transf})-(\ref{p3transf}). For the kinematics of [27], inferring
the boost parameter
from the beam velocity (which is measured accurately in the
TPA resonance absorption technique via the shifted laser frequencies),
a lower limit on $\kappa c$ of $91\: eV$ is obtained directly by
attributing the
experimental uncertainty quoted above, to the presence of the new
$\kappa$-dependent terms
modifying the special relativistic prediction.
The main factors limiting the magnitude of this lower bound include
the low energy (optical) transitions which were used, and the low
velocity of the atom beam employed ($v/c \sim 4\times 10^{-3}$).

\subsection*{Michelson-Morley Experiment}

Another fundamental special relativistic test is the Michelson-Morley
experiment. As we shall see below, in order to
$\kappa$--generalise the analysis of this experiment, we need some
consideration as
to the likely forms of the $\kappa$--deformed space-time transformations.
Due to the semigroup structure of the
$\kappa$--Lorentz
transformations [11] in momentum space, the extension to coordinate space
is most
naturally introduced in the context of relativistic mechanics using the
full eight-
dimensional phase space. If it is assumed that the canonical phase space
Poisson
bracket
			 \begin{equation}
			 	\{x^\mu, p_\nu \} = \delta^\mu_\nu,
			  \quad \{p_\mu, p_\nu \} = 0 = \{x^\mu, x^\nu \}
			 \label{Poisson}
			 \end{equation}
be preserved under the $\kappa$--Lorentz transformations [11], then the
space-time
coordinates $x^\mu$ behave as a four-vector dual to the differentials
$dp_\nu$, so
that the complete phase-space transformation rules are
             \begin{equation}
			 p'_\mu = L_\mu(\alpha;p), \quad
			 {x'}^\mu = {(L^{-1})_\rho}^\mu (x^\rho-a^\rho)
			 \label{xtransforms}
			 \end{equation}
where $\alpha$ is the rapidity parameter of the boost, $a^\mu$ describes a
four-translation, and the matrix $L^{-1}$ is the inverse of the matrix
of partial derivatives,
              \begin{equation}
              {L_\mu}^\nu \equiv \frac{\partial L_\mu(\alpha;p)}{\partial
p_\nu}.
               \label{Ldefn}
               \end{equation}
>From (\ref{xtransforms}) we can deduce the transformation properties of
classical fields in space-time from that of their realisations
in momentum space by a standard Fourier transform. Hence for the
simple case of scalar fields $\Phi(x)$, we obtain the following expression
for the
$\kappa$--transformed field:
		       	 \begin{eqnarray*}
			     \Phi'(x') &=& \int d^4x K(\alpha;x'-a,x)\Phi(x), \\
			     K(\alpha;x', x) &=& \int \frac {d^4p}{(2\pi)^4} \exp
			     (i(p_\mu x'^\mu-L_\mu(\alpha;p)x^\mu)). 				         	
		       	 \label{fieldtransform}
		       	 \end{eqnarray*}
In the special case of a plane wave,
$\Phi(x) = A \exp(ik_\mu x^\mu)$, we have
		          \begin{equation}
		           \Phi'(x') = A \left|  L\right|_{p=\hbar k}
		           \exp i( L_\mu(\alpha;\hbar k)x'^\mu /\hbar)
		           	\label{planewave}
		           \end{equation}
where the determinant $|L|$ arises as a Jacobian factor.		

\begin{figure}[tbp]
	\centerline{\epsfbox{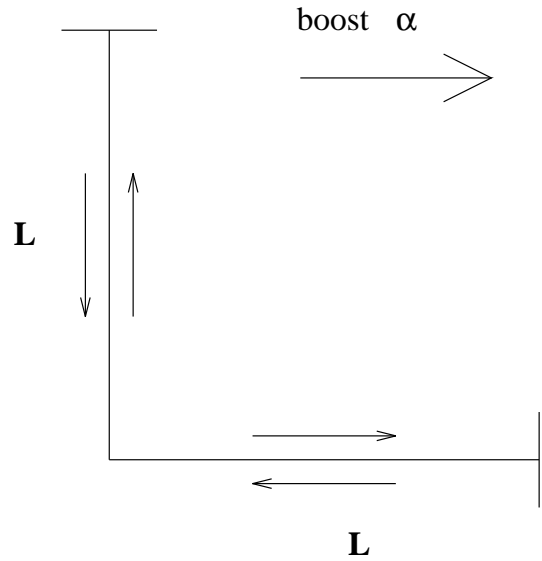}}
	\caption{Diagram of the Michelson Morley setup. The parallel
direction is taken along the $z$ axis, and the transverse direction is
taken to be along the $x$ axis. The length of each arm of the
interferometer is $L$, and the preferred frame moves with rapidity $\alpha$
with respect to the laboratory frame.}
	\protect\label{MMfig}
\end{figure}

The re-analysis of the Michelson-Morley experiment is accomplished
using a classical scalar field model. Assuming that the $\kappa$--deformed
approach is compatible with a preferred-frame theory, we wish to set up a
formalism in which monochromatic radiation in each of two orthogonal arms
of the
interferometer in the laboratory frame (described by plane waves) is
written in terms of
the $\kappa$--transformed expression for the scalar field
as seen by a moving observer (the preferred frame). Ultimately, we need to
evaluate
the $\kappa$--transformed fields at the coordinates of the recombination
point (see figure \ref{MMfig}). Unfortunately however, the expression
(\ref{planewave}) only gives these fields \emph{in the
coordinates of the moving observer}.
Thus we need to use the $\kappa$--deformed space-time transformations,
(\ref{xtransforms}). We make the additional assumption that the
four-momentum dependence inherent in these space time
transformations (\emph{c.f.} (\ref{Ldefn})) is \emph{simply the
four-momentum of the
radiation used}. The $\kappa$--Lorentz transformed plane wave scalar field,
in terms of the \emph{laboratory} coordinates therefore takes the form
\begin{equation}
	\Phi'(x'(x)) = A \left|  L\right|_{p=\hbar k}
    \exp i( L_\mu(\alpha;\hbar k){(L^{-1})_\rho}^\mu x^\rho /\hbar)
	\label{planewave2}
\end{equation}
It is the `superposition' of two such quantities, corresponding to the
appropriate field amplitudes from each of the interferometer arms,
which will determine the observed phase shift at the recombination point
(see (\ref{funnysuperpos}) below).

The exponent of (\ref{planewave2}) will be written as
$i(L_\mu \frac{\partial}{\partial L_\mu}p_\rho x^\rho /\hbar)$.
This shorthand notation for the phase indicates
a possible procedure for its evaluation, using the implicitly defined
function $p_\mu(L)$, via inversion of the relevant series expansions.
>From this point of view it is also evident how the frame independence of
the phase, and hence the usual null result for the Michelson-Morley
experiment, arises from the \emph{linearity} of the standard Lorentz
transformation.
In practice, the exponent will be evaluated using the series
expansions of the appendix, and explicitly inverting the matrix $L$ to the
appropriate order in $\kappa$.

Using the geometry of figure 1, where the $x$ axis is vertical and the
$z$ axis horizontal, the laboratory wave 4-vectors of the radiation in the
emitted,
or outward ($+$) and reflected, or inward ($-$), moving beams are
\begin{eqnarray*}
	(k^{\parallel})^\pm & = & (k_0,0,0,\pm k_3)  \\
	(k^{\perp})^\pm & = & (k_0,\pm k_1,0,0)
\end{eqnarray*}
respectively. Ignoring an inessential overall phase, and phase change
on reflection, and matching the phase of the emitted and reflected
beams at the mirror positions
\begin{eqnarray*}
	(m_{\parallel}) & = & (ct,0,0,L),  \\
	(m_{\perp}) & = & (ct,L,0,0),
\end{eqnarray*}
the total field amplitude at the recombination point $(r)$
$=(ct,0,0,0)$ (taken to be the origin in the laboratory frame) is finally
\begin{equation}
	\Phi'(r) = A'_\parallel \exp(L \frac{\partial}{\partial L}k^-_\parallel
	\cdot r + L \frac{\partial}{\partial L}(k^-_\parallel - k^+_\parallel)
	\cdot m_\parallel) +
	A'_\perp \exp(L \frac{\partial}{\partial L}k^-_\perp
	\cdot r + L \frac{\partial}{\partial L}(k^-_\perp - k^+_\perp)
	\cdot m_\perp),
	\label{funnysuperpos}
\end{equation}
where $ A'_\parallel, A'_\perp$ are the modified plane wave amplitudes,
as in (\ref{planewave2}).

Working to order $\kappa^{-2}$, the effect of the
(spacetime independent) difference between the
$A'$ factors on any resultant fringe pattern can be ignored (since it
contributes only
to the overall amplitude of the intensity modulation). The modulation
itself is determined by the phase difference $\delta\phi$ between the two terms in
(\ref{funnysuperpos}). Using the series expansions from the appendix, we
find
\begin{equation}
	\delta \phi = \frac {\hbar^2 k^3}{12 \kappa^2}
	\left[ L \sinh^2 \alpha(6+10 \sinh^2 \alpha) - ct \sinh \alpha
	(6 \cosh \alpha + 10 \cosh \alpha \sinh \alpha + 9 \sinh^3 \alpha) \right].
	\label{finaldeltaphi}
\end{equation}

As mentioned above, this non-null prediction for the Michelson-Morley
experiment appears to
imply the existence of a preferred frame.
It is only when the
interferometer is in this preferred frame that there will be found to be no
phase shift for
all orientations. If we $\kappa$--Lorentz boost to the frame of the Earth
for example,
a phase shift will be observed depending upon the orientation of the
interferometer apparatus with respect to the preferred frame.
An important feature of (\ref{finaldeltaphi}) is its time dependence.
Of course, if the phase variation is rapid compared with the time over
which intensity observations are made, then the phase difference washes
out and the experiment apparently is consistent with a null result
(leading in principle to an \emph{upper} limit for $\kappa c$).
Moreover, it can be seen from the $\sinh \alpha$ dependence
of the terms in (\ref{finaldeltaphi}) that the time variation of the phase shift
is in fact more sensitive to $\kappa$ than the dependence on the interferometer
arm length $L$.

To obtain limits on $\kappa$ from the above, consider for example the
experiment of [28]. By considering the fractional change in an etalon of
length,
the anisotropy of space was measured to be less than
$\delta L/L =\varepsilon=1.5\times 10^{-15}$, by means of a
frequency servo-stabilised laser system which tracked a particular etalon cavity
mode to within $\delta \omega / \omega = \varepsilon$ as the (single) arm
rotated
on a table at angular frequency $\delta \omega$. If it is assumed for this
case that the phase can be analysed along the above lines, then no
signature of the time dependent term $\omega_\kappa t$ in (\ref{finaldeltaphi})
would appear provided $\omega_\kappa < \delta \omega$ (a frequency variation
found at $\delta \omega$ in the experiment [28] was attributed to
gravitational flexure of the apparatus). Thus for small $\alpha$ we have
from (\ref{finaldeltaphi})
\[
\frac {({\hbar^2 k^3}/{2 \kappa^2})\alpha c }{kc}<\varepsilon
\]
leading, for the laser frequency used,
and taking the cosmic background radiation frame
($\beta = 1.3 \times 10^{-3}$) as the preferred
frame, to a lower limit on $\kappa c$ of $250 \: keV$. Clearly, a true
Michelson-Morley experiment explicitly designed to
confirm the stability of the phase to the level of $\varepsilon$ over an
interval of
several months or years would potentially provide a very stringent test
of $\kappa$.

On the other hand, the \emph{instantaneous}
phase shift (the length-dependent piece of (\ref{finaldeltaphi})) is
difficult to isolate experimentally without additional means to null the time
variation (perhaps by means of a multiple-arm apparatus).
Nevertheless, for the purposes of illustration, we can interpret
the limit $\delta L/L =\varepsilon=1.5\times 10^{-15}$
of [28] as providing a null result for an experiment of this type.
Again, for small $\alpha$ we have from (\ref{finaldeltaphi}) in this case
\[
\frac {({\hbar^2 k^3}/{4 \kappa^2}) L \alpha^2 }{k L}<\varepsilon
\]
leading to the lower
limit on  $\kappa c$ of $6.2 \: keV$.
\section{Conclusions}

In this paper we have investigated some of the important consequences
associated with the $\kappa$--deformed Poincar\'{e} group based on and
extending from the work
done by [11]. In the two major relativistic tests studied, we have obtained
lower
limits on $\kappa$, by comparing the
modified theory with experiment. The lower limits on $\kappa$
thus obtained were however found to be small, particularly when compared
with limits obtained from particle
accelerator [2] and astrophysical [22] tests.
It is perhaps surprising that two of the
standard tests, normally regarded as strong confirmation of special relativity,
turn out to be relatively insensitive to $\kappa$. The main reason is
the low energy nature of the precision tests.
In fact, any modifications to special relativity,
including the $\kappa$--deformation, are most likely to be
significant for high energy processes
only. One could for example substantially improve the lower limits obtained
on $\kappa$ by performing
the Michelson-Morley experiment using particles with a non-zero rest
mass, say neutrons. The four-momentum
component $p_0 c$ will now have a very large  rest mass  contribution
($\sim 1\: GeV$).
Assuming that an experimental accuracy of
$1.5\times 10^{-15}$ is again feasible, much larger lower limits on $\kappa$
in the $10^{3} \: GeV$ range, comparable with the limits [2] from particle
accelerator experiments, are obtainable.

Beyond relativistic tests, our approach in the paper has been to
underline \emph{in principle} various physical implications
of the modified relativity, pointing out the essential assumptions.
The transverse Doppler analysis and the Thomas precession use
only the $\kappa$-deformed Poincar\'{e} algebra, whereas both the Casimir
effect and the Michelson-Morley analysis involve
additional assumptions. In the former, additivity of the energy
of (virtual) multiphoton states has been used implicitly
in addition to the modified dispersion relation. While this seems justified
given the fact that the $P_0$ coproduct is undeformed, nevertheless
other more subtle effects to do with composite systems
in a fully developed $\kappa$--deformed Maxwell gauge theory cannot be ruled
out-- perhaps the divergence of the result (\ref {Casimireffect})
is an indication of the incompleteness of the analysis.
In the case of the Michelson-Morley experiment,
while our procedure involving momentum-dependent spacetime transformations
appears to have been invoked merely to provide a recipe
for obtaining a concrete answer for this particular case,
it is certainly to be expected that a complete theory would involve some
kind of
modification to the `superposition' principle.
Thus at the level of \emph{quantum} fields, where
intensity measurements correspond to appropriate correlation
functions, it is well possible that
the necessity for nontrivial coproducts forces the intervention of
certain grouplike operators (such as $\exp (P_0/\kappa)$). These may have the
same kind of effect on the amplitudes as has been invoked here
as an extra assumption to do with the classical fields.

In conclusion, the analysis of experimental effects presented
here can be regarded as a preliminary to a more
far-reaching study. In the absence of a definitive
and consistent deformed field theory, one more modest goal in this context
would be to develop further a $\kappa$-relativistic particle mechanics,
especially for spin. Certainly, if the
formalism can be re-cast as an effective metric theory, then
a host of experimental results, intended as tests of general relativity,
could be re-interpreted in terms of limits on the deformation parameter
$\kappa$ in the spirit of this work.

\subsection*{Acknowledgements}

The authors would like to thank
Prof Henri Ruegg for helpful suggestions in the early stages of the
work, and for drawing our attention to ref [30];
 Dr Lindsay Dodd for correspondence on the deformed Maxwell equations;
Prof Angas Hurst for critical comments; Prof Jerzy Lukierski for
supplying us with a copy of the
Domokos preprint (ref [22]), and for continuous support and encouragement,
and Prof Geoff Stedman for useful discussions. We also
thank the referees for enforcing a more focussed presentation.

\section*{Appendix}
\setcounter{equation}{0}
\renewcommand{\theequation}{A.\arabic{equation}}
The $\kappa$--deformed Poincar\'{e} algebra [11] has the following structure.
\begin{eqnarray}
\left[ M_{i},M_{j}\right]
&=&i\hbar\epsilon_{ijk}M_{k}
\;\;\;\;\; \;\;\;\;\;\;\;\;\;\;\;\;\;\;\;\;\;\;\;
\;\;\;\;\;\;\;\;\;\;\;\;\;\;\;
\left[P_\mu,P_\nu\right]=0 \nonumber \\
\left[ L_{i},M_{j}\right]&=&i\hbar\epsilon_{ijk}L_{k} \nonumber \\
\left[ M_{i},P_{j}\right]&=&i\hbar\epsilon_{ijk}P_{k}
\;\;\;\;\;\;\;\;\;\;\;\;\;\;\;\;\;\;\;\;\;\;\;\;\;\;
\;\;\;\;\;\;\;\;\;\;\;\;\;\;
\left[M_i,P_0\right]=0 \nonumber \\
\left[ L_{i},P_{j}\right]&=&i\hbar\kappa\delta_{ij}\sinh{P_{0}\over
\kappa} \;\;\;\;\;\;\;\;\;\;\;\;\;\;\;\;\;\;\;\;\;\;\;\;\;
\;\;\;\;\;\;\;
\left[L_{i},P_{0}\right]=i\hbar P_{i} \nonumber\\
\left[ L_{i},L_{j}\right]&=&-i\hbar\epsilon_{ijk}\left(
M_{k}\cosh{P_{0}\over\kappa}-{1\over4\kappa^2}P_{k}\left({\bf
P.M}\right)\right),
\end{eqnarray}
where $M_i$ is the rotation operator, $L_i$ is the Lorentz boost generator,
and $P_\mu$ is the
momentum operator, all of which are Hermitian. The deformed elements in
(A1) revert
to their normal Poincar\'{e} algebraic forms in the limit of $\kappa
\rightarrow \infty$.

The coalgebra, coproduct, and the associated counit and antipodes, of the
Hopf algebra
are given [11] by:
\begin{eqnarray}
\bigtriangleup \left(M_{i}\right)&=&M_{i}\otimes {\bf 1}+{\bf
1}\otimes M_{i} \nonumber \\
\bigtriangleup \left(L_{i}\right)&=&L_{i}\otimes
e^{P_{0}/2\kappa}+e^{-P_{0}/2\kappa} \otimes
L_{i}+\frac{1}{2\kappa}\epsilon_{ijk}
\left(P_{j}\otimes M_{k} e^{P_{0}/2\kappa}+e^{-P_{0}/2\kappa} M_{j}\otimes
P_{k}\right)
\nonumber \\
\bigtriangleup \left(P_{i}\right)&=& P_{i}\otimes
e^{P_{0}/2\kappa}+e^{-P_{0}/2\kappa}\otimes P_{i}, \;\; \;\;\;\;\;\;\;
\bigtriangleup\left(P_{0}\right)= P_{0}\otimes {\bf 1}+{\bf 1}\otimes P_{0}
\end{eqnarray}
where the counits $\epsilon$ of all generators are zero. The antipode is
\begin{equation}
S \left(M_{i} \right)=-M_{i} \textstyle{,}\;\;\;\;\;\;\; S \left(P_{\mu}
\right)=-P_{\mu}  \textstyle{,}\;\;\;\;\;\;\;  S\left (L_{i}\right
)=-L_{i}+\frac{3}{2}\frac{i}{\kappa}P_{i} \textstyle{.}
\end{equation}
It should be noted that the coproduct for the space translations, and
boosts are
deformed, as is the antipode for the boosts.

>From the $\kappa$--deformed Poincare algebra (A1) we can obtain the
corresponding
generalised Lorentz transformations in four-momentum space. This can be
achieved by
considering the orbits in $(P_0,P_3)$ space of the `one parameter subgroups'
generated by exponentiation, \emph{i.e.}
\begin{equation}
P_{\mu}(\eta)=e^{i\eta L_{3}}P_{\mu}e^{-i\eta L_{3}} \textstyle{.}
\end{equation}
By differentiating (A4) with respect to $\eta$, and using the
commutation relations (A1) we
can express the deformed algebra (A3) as the following differential equation,
\begin{equation}
\ddot{p_{0}}-\kappa\sinh\frac{p_{0}}{\kappa}=0 \textstyle{.}
\end{equation}

This equation has the same form as the differential equation of the hyperbolic
pendulum, with the `rapidity' $\eta$ playing the role of time, and with the
first integral of motion describing the constant energy surfaces. By
integrating
(A5) with respect to $p_0$
we obtain the set of $\kappa$--generalised four-momentum transformations,
(see [11]). In these transformations the hyperbolic functions
$cosh$ and $sinh$ associated with the
Lorentz transformation  are replaced by the Jacobi elliptic functions
$nc$, $sc$ and $dc$. For
example for the usual boost with rapidity $\alpha$ in the x direction,
the role of $cosh \alpha$ in the Lorentz transformation is played by,
\[
nc \left(\left(1+{\epsilon}^{2}\right)^{1/2}\alpha|(1+\epsilon^{2}
)^{-1}\right)  \quad          \mbox{\rm  where}  \quad
  \epsilon=\frac{\sqrt{c^{2}M_{0}^{2}
+p_{1}^{2}+p_{2}^{2}}}{2 \kappa}.
\]
We apply these $\kappa$-generalised four-momentum transformations to
relativistic problems
in a series expanded form. To evaluate these it is first necessary to find
suitable
expansions of the Jacobi elliptical functions. These Jacobi elliptic
functions all have the
same parameter $m=1/(1+\epsilon^2)$ which is very close to unity for
large $\kappa$, (\emph{i.e.} small $\epsilon$).
Alternatively as the complementary parameter $m_1 = 1-m$ is very small, it is useful to
express these elliptical functions in terms of a power series in $m_1$.
This involves the use [29]
of the Jacobi imaginary transformations
$nc(u|m)=cn(iu|m_{1})$,  $sc(u|m)=-isn(iu|m_{1})$, and
$dc(u|m)=dn(iu|m_{1})$,
and the following Fourier-like expansions of the Jacobi elliptical
functions in powers of the parameter $m$,
\begin{eqnarray}
cn(u|m)&=&\frac{2\pi}{m^{1/2}K}\sum_{n=0}^{\infty}\frac{q^{n+1/2}}{1+q^{2n+1
}}\cos(2n+1)\nu ,\nonumber \\
sn(u|m)&=&\frac{2\pi}{m^{1/2}K}\sum_{n=0}^{\infty}\frac{q^{n+1/2}}{1-q^{2n+1}}
\sin(2n+1)\nu ,  \nonumber \\
dn(u|m)&=&\frac{\pi}{2K}+\frac{2\pi}{K}\sum_{n=1}^{\infty}\frac{q^{n}}{1+q^{
2n}}\cos 2n\nu,
\end{eqnarray}
where the nome $q=\exp(-\pi K'/K)$ and the argument $\nu = \pi u/2K$ are
defined in terms of the real and imaginary quarter periods $K, K'$,
both of which are expressible in series expansion form.

Finally, the series expansions of the generalised transformations are found
to take the following forms,
\begin{eqnarray}
p_{0}'& = &p_{0}\cosh\alpha +p_{3}\sinh\alpha +\frac{1}{4\kappa^{2}}\left(\frac{
-\left(p_{3}\sinh\alpha
+p_{0}\cosh\alpha\right)^{3}}{6}+\frac{p_{0}^{3}\cosh\alpha}{6}\right)
 \nonumber \\ &+&
\frac{p_{3}\mu^{2}}{4\kappa^{2}}\left(\frac{\alpha\cosh\alpha}{2}-\frac{\sin
h\alpha}{2}+\frac{p_{0}
^{2}\sinh\alpha}{2\mu^{2}}+\frac{\sinh^{3}\alpha}{2}+\frac{p_{3}^{2}\sinh^{3}
\alpha}{\mu^{2}}+\frac{\cosh\alpha\left(\sinh2\alpha-2\alpha\right)}{8}\right)
 \nonumber \\
&+&\frac{p_{0}\mu^{2}}{4\kappa^{2}}\left(\frac{\alpha\sinh\alpha}{2}+\frac{p
_{3}^{2}\cosh\alpha\sinh
^{2}\alpha}{\mu^{2}}+\frac{\sinh\alpha\left(\sinh
2\alpha-2\alpha\right)}{8}\right)
+0(\kappa^{-4})
\label{p0transf}
\end{eqnarray}
\begin{eqnarray}
 p_{3}'& =&  p_{3}\cosh\alpha+p_{0}\sinh\alpha \nonumber \\
&+&\frac{1}{4\kappa^{2}}\left(\frac{p_{0}
^{3}\sinh\alpha}{6}+\frac{p_{3}\sinh\alpha (2\alpha\mu^{2}+3\mu^{2}\sinh
2\alpha+4p_{3}^{2}\sinh 2\alpha)}{8}\right) \nonumber \\
&+&
\frac{p_{0}\mu^{2}}{4\kappa^{2}}\left(\frac{\alpha\cosh\alpha}{2}-\frac{\sin
h\alpha}{2}+
\frac{p_{0}^{2}\sinh\alpha}{2\mu^{2}}+\frac{p_{3}^{2}\sinh^{3}\alpha}{\mu^{2}}
+\frac{\cosh\alpha(\sinh2\alpha-2\alpha)}{8}\right) \nonumber \\
&+&0(\kappa^{-4})
\label{p3transf}
\end{eqnarray}
where     $\mu=\sqrt{c^{2}M_{0}^{2}+\pi_{1}^{2}+\pi_{2}^{2}}$, and $\pi_i
\equiv p_i$ for the transverse components.

\subsection*{References}
\noindent
[1]	Drinfeld W G Proc International Congress of Mathematics (Berkeley,
USA,1986) p 70\\
\noindent
[2]	Ruegg H 1993 $q$-deformation of semisimple and non semi simple Lie
algebra's
In: NATO ASI  Series C: Mathematical and Physical Sciences Vol 409 ed L A
Ibort and M A Rodriguez (Kluver, Dordrecht, 1993) p 45-82.\\
\noindent
[3]	Lukierski J, Nowicki A and Ruegg H 1991 Phys. Lett. 271B 321\\
\noindent
[4]	Maslanka P 1993 J. Math. Phys. 34 6025\\
\noindent
[5]	Lukierski J and Ruegg H 1994 Phys. Lett. 329B 189\\
\noindent
[6]	Ballesteros A, Herranz F J, del Olmo M A and Santander M 1993 Quantum
algebras for maximal motion groups of N-dimensional flat spaces DFT Valladolid
preprint, hep-th/9404052\\
\noindent
[7]	Dobrev V, Lukierski J, Sobczyk J and Tolstoy V N 1993 preprint ICTP Trieste
IC/92/188\\
\noindent
[8]	Lukierski J, Nowicki A and Sobczyk J 1993 J. Phys. A26 L1099\\
\noindent
[9]	Kosinski P, Lukierski J, Maslanka P and Sobczyk J 1994 Quantum
deformation of
the Poincare supergroup and $\kappa$-deformed superspace. Wroclaw preprint
IFT UWr
868/94\\
\noindent
[10]	Lukierski J, Nowicki A and Ruegg H 1992 Phys. Lett. 293B 344\\
\noindent
[11]	Lukierski J, Ruegg H and Ruhl W 1993 Phys. Lett. 313B 357\\
\noindent
[12]	Biedenharn L C, Mueller B and Tarlini M 1993 Phys. Lett. 318B 613\\
\noindent
[13]	Giller S, Kunz J, Kosinski P, Majewski M and Maslanka P 1992 Phys.
Lett. 286B
57\\
\noindent
[14]	Ruegg H and Tolstoy V N 1994 Lett. Math. Phys. 32 85\\
\noindent
[15]	Zumino B 1994 Beyond supersymmetry; calculus on quantum spaces and
quantum groups invited talk XXth ICTGMP Osaka 1994/07-828\\
\noindent
[16]	Podles P and Woronowicz S L 1990 Commun. Math. Phys. 130 381\\
\noindent
[17]	Majid S 1988 Class. Quantum Grav. 5 1587\\
\noindent
[18]	Majid S 1990 Pacific J. Math. 141 311\\
\noindent
[19]	Majid S 1994 Duality principle and braided geometry Cambridge preprint
DAMPT/94-69, Proc 1st Gursey Memorial conference on strings and symmetries\\
Istanbul
\noindent
[20]	Majid S and Ruegg H 1994 Phys. Lett. 334B 348\\
\noindent
[21]	Hellings R W 1984 General relativity and gravitation ed B Bertotti F
de Felice
and A Pascolini (Reidel  Dordrecht) p 365\\
\noindent
[22]    Domokos G and Kovesi-Domokos S 1994 J Phys G 20 1989 
\\
\noindent
[23]	Ballentine L E 1990 Quantum mechanics (Prentice Hall)\\
\noindent
[24]	Milonni P W and Shih M L 1992 Contemp. Phys. 33 313\\
\noindent
[25]	Sparnaay M J 1958 Physica 24  751\\
\noindent
[26]	Bowes J P 1994 Honours Thesis University of  Tasmania\\
\noindent
[27]	Kaivola M, Poulsen O, Riis E, and Lee S 1985 Phys. Rev. Lett. 54
255\\
\noindent
[28]	Brillet A and Hall J L 1979 Phys. Rev. Lett. 42  549\\
\noindent
[29]	Abramowitz M and Segun I 1968 Handbook of Mathematical Functions
(Dover).\\
\noindent
[30]    H Ruegg, C Gomez and Ph Zaugg 1994 J Phys A 27 7805
\newpage
\noindent

\end{document}